\documentclass[12pt]{article}

\catcode`\@=11
\@addtoreset{equation}{section}

\evensidemargin \oddsidemargin

\usepackage[dvipdfmx]{graphicx,hyperref}
\hypersetup{
setpagesize=false,
bookmarksnumbered=true,
bookmarksopen=true,
colorlinks=true,
linkcolor=black,
citecolor=black,
urlcolor=black,}
\usepackage{mathrsfs,amsbsy,amssymb,latexsym,amsfonts,amsmath,
amssymb,
}

 \usepackage{bm}

\usepackage[dvipdfmx]{graphicx}

\newcommand\CL{\mathcal{L}}

\newcommand\CH{\mathcal{H}}

\newcommand\pa{\partial}

\newcommand\bbE{\mathbb{E}}

\newcommand\nn{\nonumber}

\newcommand\adss[2]{AdS$_{#1}\times$S$^{#2}$}

\newcommand\e{{\rm e}}

\renewcommand\d{{\rm d}}

\newcommand\bref[1]{(\ref{#1})}

\begin{document}

\vspace*{3cm}

\begin{center}
{\Large \bf 
Non-commutative M-branes from\\ Open
 Pure Spinor  Supermembrane 
}
\vspace*{3cm}\\
{\large Sota Hanazawa\footnote{\texttt{16nd109n@vc.ibaraki.ac.jp}}
and
Makoto Sakaguchi\footnote{\texttt{makoto.sakaguchi.phys@vc.ibaraki.ac.jp}}
}
\end{center}
\vspace*{1.5cm}
\begin{center}
$^*$
Graduate School of Science and Engineering, Ibaraki University, Mito 310-8512, Japan

$^\dag$ Department of Physics, Ibaraki University, Mito 310-8512, Japan
\end{center}

\vspace{2cm}

\begin{abstract}

Open supermembrane with a constant three-form
flux in the pure spinor formalism is examined.
The BRST symmetry of the open supermembrane action
leads to non-commutative (NC) M-branes.
In addition to the NC M5-brane with a self-dual two-form flux,
we find a NC M9-brane with an electric flux 
and a NC M9-brane with a magnetic flux.
The former reduces in the critical electric flux limit to an M2-brane on the M9-brane,
while the latter reduces in the strong magnetic flux limit to infinitely many Kaluza-Klein monopoles
dissolved into the M9-brane.
These NC M-branes
are shown to preserve  a half of 32 supersymmetries.

\end{abstract}

\thispagestyle{empty}
\setcounter{page}{0}

\newpage

\tableofcontents

\section{Introduction}

Supermembrane theory in eleven-dimensions\,\cite{smem}
 is expected to be closely related to
a formulation of M-theory\,\cite{M-theory},
despite the fact that its microscopic degrees of freedom
have not been completely understood yet.
Fortunately at the semiclassical level,
M-theory is an eleven-dimensional supergravity theory interacting with
1/2 BPS objects:
M2-branes, M5-branes, M9-branes, M-waves and Kaluza-Klein (KK) monopoles.
It is known  that an open supermembrane can end on Dirichlet $p$-branes with $p=1,5$ and $9$ \cite{EMM}\cite{dWPP}.
The $p=5$ case corresponds to the M5-brane
and the $p=9$ case to the M9-brane
which is the boundary of the eleven-dimensional spacetime in
the Ho\v rava-Witten formulation of the heterotic string theory \cite{HW}.
In \cite{KK7M}, 
the M-wave \cite{M-wave} 
of eleven-dimensional supergravity
 are identified with D0-branes of IIA supergravity
 and their dual D6-branes are shown to be the 
KK monopole\,\cite{KK monopole} 
in eleven-dimensions.

Dirichlet branes of a $\kappa$-symmetric open supermembrane
are investigated from the $\kappa$-symmetry argument\,\cite{EMM}.
Furthermore, non-commutative (NC) M-branes
are discussed from $\kappa$-symmetry
of a $\kappa$-symmetric open supermembrane with
a constant three-form flux
and it is found that 
the self-duality of the two-form gauge field on the M5-brane world-volume
follows from $\kappa$-symmetry of the open supermembrane\,\cite{SY NC M5}.
In addition, intersecting NC M-branes are discussed in \cite{SY intersecting,SY NC M5 2}.

\medskip

In this paper we examine an open supermembrane with a constant three-form
flux in the pure spinor formalism\,\cite{PS mem}.
In the pure spinor formalism, the $\kappa$-symmetry is replaced with the BRST
symmetry.
One of advantages of our approach is to consider BRST symmetry
which is expected to survive quantum corrections.
It implies that our analysis in this paper may give a quantum consistency check
for  the $\kappa$-symmetry arguments.
In addition, we will derive a NC M9-brane with an electric flux 
and a NC M9-brane with a magnetic flux.
The former reduces in the critical electric flux limit to an M2-brane on the M9-brane.
The latter reduces in the strong magnetic flux limit to infinitely many KK monopoles dissolved into the M9-brane,
and is identified with a bound state of an M9-brane and KK monopoles.

We will examine the supersymmetry variation of an open supermembrane
in the pure spinor formalism,
and find that the boundary condition for the BRST symmetry
solves those for supersymmetry.
This shows that the NC M-branes derived in this paper 
preserve a half of 32 supersymmetries
and should be 1/2 BPS objects.

\medskip

This paper is organized as follows.
In the next section, we introduce an open supermembrane action with 
constant fluxes in the pure spinor formalism,
and give the BRST transformation law and the supersymmetry transformation law
under which the action is invariant.
We derive surface terms of the BRST transformation
and deduce boundary conditions to eliminate them in  section 3.
In section 4,
as solutions of the boundary conditions,
we obtain NC M-branes;
a NC M5-brane with self-dual three-form fluxes,  
a NC M9-brane with an electric flux and a NC M9-brane with a magnetic flux.
In addition, the NC M-branes are shown to be half supersymmetric,
namely 1/2 BPS.
The last section is devoted to summary and discussions.
In the appendix A, we give a derivation of  NC M5-branes,
and we describe a derivation of surface terms for the supersymmetry transformation
of the action  in the appendix B.

\section{Pure Spinor Supermembrane  with constant three-form fluxes} 

Before introducing the pure spinor supermembrane action, we
introduce 
the $\kappa$-symmetric supermembrane action in $d=11$.
It is composed of two parts
\cite{smem}
\begin{align}
S=&\int_\Sigma \d^3 \tau \left( \CL_0+\CL_\mathrm{WZ} \right) ~,
\label{M2}
\\
\CL_0=&
P_\mu \Pi^\mu_0
+e^0(P^\mu P_\mu +\det(\Pi^\mu_I \Pi_{J\mu}))
+e^I P_\mu \Pi^\mu_I
~,
\\
\CL_\mathrm{WZ}= &\,\epsilon^{ijk}\Bigg[
\frac{1}{6}\CH_{\mu\nu\rho}
\pa_i x^\mu\pa_j x^\nu\pa_k x^\rho
-\frac{i}{4}\Big\{
\bar\theta \Gamma_{\mu\nu} \pa_i\theta\, \pa_j x^\mu  \pa_k x^\nu
\nn\\
&~~~~~
+\frac{i}{2}\bar\theta \Gamma_{\mu\nu}\pa_i\theta\,\bar\theta \Gamma^{\mu}\pa_j\theta\, \pa_kx^\nu
-\frac{1}{12}\bar\theta\Gamma_{\mu\nu}\pa_i\theta\, \bar\theta\Gamma^\mu \pa_j\theta\, 
\bar\theta\Gamma^\nu \pa_k\theta
\Big\}
\Bigg]
~,
\label{WZ}
\end{align}
where $x^\mu\, (\mu=0,1,\cdots,9,10)$ are flat spacetime coordinates,
and 
$\tau^i$  ($i=0,1,2$, and $I=1,2$ are space indices) are coordinates on the  world-volume $\Sigma$.
We have introduced $\Pi^\mu_i\equiv \pa_i x^\mu +\frac{i}{2}\bar\theta \Gamma^\mu \pa_i\theta$.
The $\Gamma^\mu$ denote $32\times 32$ gamma matrices
and $\theta^\alpha$ is a 32-component Majorana spinor in $d=11$.
We define  $\bar \theta$ by $\bar \theta=\theta^{\rm T} C$ with the charge conjugation matrix $C$
so that $(C\Gamma^{\mu_1\cdots \mu_n})_{\alpha\beta}$ is symmetric under the exchange 
$\alpha\leftrightarrow \beta$
iff $n=1,2$ mod 4.
The $e^0$ and $e^I$ are Lagrange multipliers for reparametrization constraints.
By eliminating $P_\mu$ by its equation of motion, $\CL_0$ reduces to
 the Nambu-Goto Lagrangian 
$\sqrt{-\det \Pi_i^\mu\Pi_{j\mu}}$
\cite{smem Hamiltonian}.

We have introduced 
a constant three-form flux
$\CH=C-{\d}b$
in $\CL_\mathrm{WZ}$
where $C$ and $b$  denote the three-form gauge potential
and the  two-form gauge field on the boundary brane, respectively.
It should be noted that the action \bref{M2} is $\kappa$-symmetric and supersymmetric
even in the presence of $\CH$.

In the pure spinor formalism, 
$\kappa$-symmetry is replaced with BRST symmetry.
The  supermembrane action in the pure spinor formalism\,\cite{PS mem}
is given as
\begin{align}
S_\mathrm{pure}=\int_\Sigma  \d^3\tau
\Big[&
\tilde P_\mu \Pi^\mu_0
+\CL_\mathrm{WZ}
+d_\alpha \pa_0\theta^\alpha
+w_\alpha\pa_0\lambda^\alpha+(d\Gamma_\mu \pa_I\theta)\,\Pi^\mu_J \epsilon^{IJ}
\nn\\&
-\frac{1}{2}(\tilde P^\mu \tilde P_\mu +\det(\Pi_I\Pi_J))
+e^I(\tilde P_\mu\Pi^\mu_I +d \pa_I\theta+w\pa_I\lambda)
\nn\\&
+(w\Gamma_\mu \pa_I\lambda)\,\Pi^\mu_J\epsilon^{IJ}
-i\epsilon^{IJ}(w\Gamma_\mu \pa_I\theta)(\bar\lambda \Gamma^\mu \pa_J\theta)
+i\epsilon^{IJ}(w\pa_I\theta)(\bar\lambda \pa_J\theta)
\Big]~,
\label{action PS}
\end{align}
where
$\tilde P_\mu$ denotes 
$\tilde P_\mu\equiv P_\mu-\frac{1}{2} B_{\mu MN}\pa_I Z^M \pa_J Z^N \epsilon^{IJ}$
with $Z^M=(x^\mu,\theta^\alpha)$.
The $B_{MNP}$ is defined by
$\CL_\mathrm{WZ}\equiv \frac{1}{6}\epsilon^{ijk} B_{MNP}\pa_i Z^M\pa_j Z^N\pa_k Z^P$
where $\CL_\mathrm{WZ}$ is given in \bref{WZ}.
Define the momentum conjugate to $\theta^\alpha$ by
\footnote{The derivative with a  superscript $r$ denotes  the right derivative.
} $p_\alpha\equiv \frac{\pa^r L}{\pa \dot\theta^\alpha}$,
and then we introduce
 $d_\alpha$
as 32 fermionic constraints
\begin{align}
d_\alpha= p_\alpha -\frac{i}{2}\tilde P^\mu(C\Gamma_\mu\theta)_\alpha
-\frac{1}{2} \epsilon^{IJ} B_{MN\alpha} \pa_IZ^M\pa_JZ^N~.
\end{align}
The Grassmann-even spinor fields $(\lambda^\alpha,w_\alpha)$ are pure spinor ghosts.
The Lagrange multiplier $e^0$ has been set to $-1/2$.

The supermembrane action \bref{action PS} is invariant under the supersymmetry transformations
\begin{align}
\delta_\epsilon \theta ^\alpha = \epsilon^\alpha~,~
\delta_\epsilon x^\mu=\frac{i}{2} \bar\theta\Gamma^\mu \epsilon~,~
\delta_\epsilon e^0=\delta_\epsilon e^I=0~,~
\delta_\epsilon\lambda ^\alpha=\delta_\epsilon w _\alpha=0~,~
\delta_\epsilon \tilde P_\mu=\delta_\epsilon d_\alpha=0~,
\label{SUSY}
\end{align}
where $\tilde P_\mu$ and $d_\alpha$ are defined to be invariant
under supersymmetry transformations.

The BRST operator\footnote{
The double spinor formalism \cite{double spinor} sheds some light on a derivation of the BRST operator
for the pure spinor supermembrane.
} 
is defined by $Q\equiv \lambda^\alpha d_\alpha$
which acts on a field $f$ by $Qf=i\{Q,f ]$.
By using (anti-)commutation relations
$\{p_\alpha,\theta^\beta\}=-i\delta^\beta_\alpha$,
$[P_\mu,x^\nu]=i\delta^\nu_\mu$
and $[\lambda^\alpha,w_\beta]=-i\delta_\beta^\alpha$,
we may derive
\begin{align}
Q\theta^\alpha=&\lambda^\alpha~,~~~
Q x^\mu=\frac{i}{2} \bar\lambda \Gamma^\mu \theta~,~~~
Q d_\alpha=-i\hat \Pi^\mu_0 (C\Gamma_\mu\lambda)_\alpha
+\frac{i}{2} \epsilon^{IJ} \Pi_{I}^\mu\Pi_J^\nu (C\Gamma_{\mu\nu}\lambda)_\alpha~,~~~
\nn\\
Q\lambda^\alpha=&0~,~~~
Qw_\alpha=d_\alpha~,~~~
{Q \tilde P_\mu=}{-i  \bar \lambda \Gamma_{\mu\nu}\pa_I \theta \,\Pi^\nu_J \epsilon^{IJ}}~.
\label{BRS}
\end{align}
For the nilpotency of the BRST operator, we impose
the pure spinor constraint $\bar\lambda\Gamma^\mu \lambda=0$, 
and its secondary constraints\footnote{
For further discussion on this issue see \cite{PS mem}\cite{11d PS constraint}.
}
$(\bar\lambda \Gamma_{\mu\nu} \lambda) \,\Pi^\mu_I=0$
and $\bar\lambda \pa_J \lambda=0$.
The equation of motion for $P_\mu$
determines $\tilde P^\mu$ as $\tilde P^\mu=\Pi^\mu_0 + e^I \Pi_I^\mu \equiv \hat \Pi^\mu_0$.
The last equation in \bref{BRS}
follows from the equation of motion
$\nabla\theta\equiv \pa_0\theta +e^I\pa_I\theta -2 e^0 \Gamma_\rho \pa_K\theta\,\Pi^\rho_L\epsilon^{KL}
=0$
with $e^0=-1/2$.
We assume that 
$
Q e^I=-i\epsilon^{IJ} \bar\lambda \pa_J \theta$,
which is suggested by the $\kappa$-transformation of $e^I$
\cite{PS mem}.

The BRST invariance of $S_\mathrm{pure}$ can be shown \cite{PS mem} 
by following the method used in \cite{OT}.
First, we note that 
\begin{align}
 S_\mathrm{pure} - \tilde S=\int \d^3\tau \,Q [w\nabla \theta]~,
\end{align}
where $\tilde S$ denotes  $S$ in \bref{M2} with the replacements $P_\mu\to \tilde P_\mu$
and $e^0\to-1/2$.
It is convenient for us to introduce a Grassmann-odd parameter $\varepsilon$ to the BRST transformations:
$\delta f=\varepsilon Q f$, namely
\begin{align}
 \delta\theta^\alpha
 =\varepsilon \lambda^\alpha~,~~~
 \delta x^\mu=\frac{i}{2} \varepsilon\bar \lambda  \Gamma^\mu  \theta~,~~~
 \delta e^I=-i\epsilon^{IJ} \varepsilon \bar\lambda \pa_J \theta~,~~~
 \delta \tilde P_\mu=-i \varepsilon \bar \lambda \Gamma_{\mu\nu}\pa_I \theta \,\Pi^\nu_J \epsilon^{IJ}~.
\label{BRST delta}
\end{align}
As in \cite{PS mem}, one may show that
\begin{align}
\delta \tilde S=&i\int\d^3\tau\,
\varepsilon\bar\lambda\left(
\Gamma^\mu \tilde P_\mu -\frac{1}{2}\Gamma_{\mu\nu} \,\epsilon^{IJ}\Pi^\mu_I\Pi^\nu_J
\right)\nabla\theta~,
\end{align}
and that
\begin{align}
\delta\int \d^3\tau \,Q [w\nabla \theta]=&
-i\int \d^3\tau \,\varepsilon \bar\lambda \left(
\Gamma_\mu \hat \Pi_0^\mu -\frac{1}{2}\Gamma_{\mu\nu}\epsilon^{IJ}\Pi_I^\mu \Pi_J^\nu
\right)\nabla \theta~.
\end{align}
Gathering these together, we may conclude that $S_\mathrm{pure}$ is BRST invariant
\begin{align}
\delta S_\mathrm{pure}=\delta \tilde S+\delta\int \d^3\tau \,Q [w\nabla \theta]
=0~.
\end{align}

\section{Open Supermembrane and BRST surface terms}

For an open supermembrane
in the pure spinor formalism,
the BRST transformation of the action leaves surface terms.
We will show that the boundary condition to eliminate them
leads to a classification of Dirichlet-branes of an open supermembrane.
Furthermore  we note that 
an open supermembrane
with constant three-form fluxes
may attach to non-commutative M-branes.

In this section, we will derive
surface terms of the BRST transformation of
the open pure spinor supermembrane  action.
First of all,
we consider boundary conditions  for bosonic variables.
The bosonic part of the pure spinor supermembrane action \bref{action PS}
is the same as the bosonic part of the action \bref{M2}
with $e^0=-1/2$. 
In studying the bosonic part, we will restore $e^0$ as a Lagrange multiplier,
and we consider the bosonic part of \bref{M2} 
which is classically equivalent to the 
Nambu-Goto action\,\cite{smem Hamiltonian}.
Varying it with respect to $x^\mu$,
we obtain the surface term
\begin{align}
\delta S_\mathrm{pure}^{bos}|
=&
\int_{\pa\Sigma}
\left(
\pa^i x_\mu
+\frac{1}{2}\epsilon^{ijk} \CH_{\mu\nu\rho}
\pa_j x^\nu \pa_k x^\rho
\right)\delta x^\mu
n_i \d S~,
\end{align}
where $n_i$ is the unit vector normal to $\pa\Sigma$.

We will turn on a constant $\CH$ along the world-volume
of the Dirichlet $p$-brane of the open supermembrane.
In order to eliminate the surface term,
we must impose either of  the following boundary conditions\,\cite{M BC}:
the
Neumann boundary condition
$\pa_\mathrm{n} x^{\bar{\mu}}
+\CH^{\bar{\mu}}
{}_{
\bar{\nu}
\bar{\rho}
}
\pa_\mathrm{t}x^{\bar{\nu}}\pa_\tau x^{\bar{\rho}}
=0$
for 
Neumann directions $x^{\bar{\mu}_a}$
($a=0,1,\cdots,p$),
or
the
Dirichlet boundary condition
$\pa_\mathrm{t} x^{\underline{\mu}}=\pa_\tau x^{\underline{\mu}}=0$
for Dirichlet directions $x^{\underline{\mu}{}_a}$
($a=p+1,\cdots,10$).
We have defined
$\pa_\mathrm{n}\equiv n^i \pa_i$
and $\pa_\mathrm{t}\equiv t^i \pa_i$
with $t$ and $\tau$ being vectors tangent to $\pa\Sigma$.
The Neumann boundary condition above 
reduces to the ordinary Neumann boundary condition
when $\CH
=0$,
while it mixes Neumann and Dirichlet boundary conditions  for $\CH\neq 0$.

\subsection{BRST surface terms}

Now we shall consider
the surface terms of the BRST transformation.
For the BRST symmetry to be unbroken in the presence of the boundary,
these surface terms must be eliminated
by appropriate boundary conditions on the fermionic variables. 

We find that the surface terms of the BRST transformation of $S_\mathrm{pure}$
 come from the WZ term
and take the form
\begin{align}
\delta S_\mathrm{pure}|
=&
\int _{\pa\Sigma}
(\CL^{(2)}
+\CL^{(4)}
+\CL^{(6)}
) \d S~,\label{L}\\
\CL^{(2)}=&
-\frac{i}{4}\epsilon^{ijk}
(\CH_{\mu\nu\rho}\,\bar\theta \Gamma^\rho \xi +\bar\theta\Gamma_{\mu\nu}\xi)
\,\pa_j x^\mu \pa_k x^\nu
\, n_i
~,
\label{L2}
\\
\CL^{(4)}=&
\frac{1}{8}\epsilon^{ijk}
\left(
\bar\theta \Gamma_{\mu\nu}\xi\cdot\bar\theta \Gamma^\mu \pa_j\theta
+\bar\theta \Gamma_{\mu\nu}\pa_j \theta\cdot \bar\theta \Gamma^\mu \xi
\right)\,\pa_k x^\nu\, n_i~,
\label{L4}
\\
\CL^{(6)}=&
\frac{i}{48}
\epsilon^{ijk}
\left(
\bar\theta \Gamma_{\mu\nu}\xi\cdot 
\bar\theta \Gamma^\mu \pa_j\theta
+
\bar\theta\Gamma_{\mu\nu}\pa_j\theta\cdot 
\bar\theta\Gamma^{\mu}\xi
\right)\,
\bar\theta \Gamma^\nu \pa_k \theta \,n_i~,
\label{L6}
\end{align}
where we have introduced $\xi$ by $\xi\equiv  \varepsilon \lambda$,
and $\CL^{(n)}$ denotes the $n$-th order terms in  $\xi$ as well as $\theta$.
It is worth noting that if we set $\xi=\delta_\kappa \theta$,
the above surface term coincides with that of the $\kappa$-symmetry
transformation
of the $\kappa$-symmetric open supermembrane action examined in 
\cite{EMM, SY NC M5, SY intersecting, SY NC M5 2}.
For the present paper to be self-contained,
we will derive conditions
on $\theta$ and $\xi$ for the surface terms to be deleted.

First we will show that $\CL^{(6)}$
in \bref{L6} vanishes due to the Fierz identity
\begin{eqnarray}
(C\Gamma_{\mu\nu})_{(\alpha\beta}(C\Gamma^{\nu})_{\gamma\delta)}=0~.
\label{Fierz}
\end{eqnarray}
One finds that
\begin{align}
\CL^{(6)}=&
-\frac{i}{24}
\epsilon^{ijk}
\bar\theta\Gamma_{\mu\nu}\pa_j\theta\cdot 
\bar\theta \Gamma^\mu \pa_k \theta\cdot 
\bar\theta\Gamma^{\nu}\xi
\,n_i
\nn\\
=&
-\frac{i}{24}
\epsilon^{ijk}
\bar\theta\Gamma_{\mu\nu}\pa_k\theta\cdot 
\bar\theta \Gamma^\mu \pa_j \theta\cdot 
\bar\theta\Gamma^{\nu}\xi
\,n_i
\nn\\
=&0~.
\end{align}
In the first equality
the Fierz identity
$\bar\theta \Gamma_{\mu\nu}\xi\cdot 
\bar\theta \Gamma^\nu \pa_j\theta
-\bar\theta \Gamma_{\mu\nu}\pa_j\theta\cdot 
\bar\theta \Gamma^\nu\xi =0
$
has been used,
and
in the
second equality
we have used the Fierz identity
$\bar\theta\Gamma_{\mu\nu}\pa_j\theta\cdot 
\bar\theta \Gamma^\mu \pa_k \theta
-
\bar\theta\Gamma_{\mu\nu}\pa_k\theta\cdot 
\bar\theta \Gamma^\mu \pa_j \theta
=0
$.
The third equality follows from
the anti-symmetricity of three indices
of $\epsilon^{ijk}$.

Next we consider 
$\CL^{(2)}$ in \bref{L2}.
The bosonic boundary condition implies that
$
\CL^{(2)}=
-\frac{i}{2}(\CH_{\bar\mu\bar\nu\bar\rho}\, \bar\theta \Gamma^{\bar\rho}\xi
+\bar\theta\Gamma_{\bar\mu\bar\nu}\xi)\pa_\mathrm{t}x^{\bar\mu}\pa_\tau x^{\bar\nu},
$
and then we require that
\begin{align}
\CH_{\bar\mu\bar\nu\bar\rho}\,
\bar\theta \Gamma^{\bar\rho}\xi
+\bar\theta\Gamma_{\bar\mu\bar\nu}\xi=0~.
\label{BC1}
\end{align}
We demand that
the boundary condition on $\xi$
is the same as that on $\theta$.
This is because the BRST transformation \bref{BRST delta} 
relates them each other as $\delta \theta=\xi$.
It implies that the BRST symmetry is preserved even in the presence of the boundary.
Before solving the boundary condition \bref{BC1},
we consider $\CL^{(4)}$ in \bref{L4}.
The bosonic boundary condition
reduces it to
\begin{align}
\CL^{(4)}=\,&
\frac{1}{4}
(\bar\theta \Gamma_{\bar\mu{\bar\nu}}\xi\cdot\bar\theta \Gamma^{\bar\mu} \pa_t\theta
+\bar\theta \Gamma_{\bar\mu{\bar\nu}}\pa_t \theta\cdot \bar\theta \Gamma^{\bar\mu} \xi
)
\pa_\tau x^{\bar\nu} 
-(t\leftrightarrow \tau)
\nn\\\,&
+
\frac{1}{4}
(\bar\theta \Gamma_{{\underline\mu}{\bar\nu}}\xi\cdot\bar\theta \Gamma^{\underline\mu} \pa_t\theta
+\bar\theta \Gamma_{{\underline\mu}{\bar\nu}}\pa_t \theta\cdot \bar\theta \Gamma^{\underline\mu} \xi
)
\pa_\tau x^{\bar\nu}  -(t\leftrightarrow \tau)
~.
\end{align}
The relation \bref{BC1} makes the first line
of the form
$\frac{1}{4}
\CH_{\bar\mu\bar\nu\bar\rho }(\bar\theta \Gamma^{\bar\rho}\xi\cdot\bar\theta \Gamma^{\bar\mu} \pa_t\theta
+\bar\theta \Gamma^{\bar\rho}\pa_t \theta\cdot \bar\theta \Gamma^{\bar\mu} \xi
)
\pa_\tau x^{\bar\nu} 
-(t\leftrightarrow \tau)$,
which vanishes due to the anti-symmetricity of the three indices of $\CH_{\bar\mu\bar\nu\bar\rho }$.
As a result, for $\CL^{(4)}=0$ we require
\begin{eqnarray}
\bar\theta \Gamma_{{\underline\mu}{\bar\nu}}\xi=0~~\mbox{or}~~~
\bar\theta \Gamma^{\underline\mu} \xi=0~.
\label{BC2}
\end{eqnarray}

Summarizing, the surface terms should disappear if \bref{BC1} and
either of two equations in \bref{BC2}
are satisfied.

\section{Non-commutative M-branes}

In this section, we will fix the fermionic boundary conditions
which solve \bref{BC1} and \bref{BC2}.
We shall impose the same boundary condition on $\theta$ and $\xi$
\begin{eqnarray}
\theta=M\theta~,~~~\xi=M\xi~,
\end{eqnarray}
where $M$ is the gluing matrix.
This is because
the BRST transformation relates them each other as $\delta \theta=\xi$.
In addition to the NC M5-brane obtained in \cite{SY NC M5},
two kinds of NC M9-branes will be presented below.

\subsection{Non-commutative M5-brane}\label{sec:NC M5}

Here we present the boundary condition for a NC M5-brane  which solve \bref{BC1} and \bref{BC2}.
Because
the derivation of the boundary condition is similar to those given in \cite{SY NC M5},
we put it in the appendix A.

A gluing matrix and fluxes for a NC M5-brane\footnote{
The gluing matrix $M=\e^{\varphi\Gamma^{012}}\Gamma^{01\cdots5}$
with fluxes $\CH_{345}=\sinh\varphi$ and $\CH^{012}=\tanh\varphi$
gives a different parametrization of the NC M5-brane above. 
} are
$M=\e^{\varphi\Gamma^{345}}\Gamma^{01\cdots5}$,
and $\CH_{012}=\sin\varphi$ and $\CH^{345}=\tan\varphi$.
As explained in the appendix A, the fluxes satisfy
\begin{eqnarray}
\frac{1}{(\CH_{012
})^2}-\frac{1}{(\CH^{345
} )^2}=
1~,
\end{eqnarray}
which is nothing but the self-duality condition\,\cite{SD}
for the two-form gauge field on the M5-brane world-volume\,\cite{SD2}.
We note that the self-duality condition for the two-form gauge fields
has been derived from the BRST symmetry
of an open supermembrane.

For $\varphi=0$, it represents a commutative M5-brane
as $M=\Gamma^{01\cdots5}$ and $\CH_{012}=\CH^{345}=0$.
On the other hand for $\varphi\to\pi/2$, 
the gluing matrix reduces to $M\to\Gamma^{012}$,
while fluxes are $\CH_{012}\to1$ and $\CH^{345}\to \infty$.
It seems that this may describe an M2-brane with 
a critical flux $\CH_{012}=1$.
However this limit is nothing but the OM limit\,\cite{OM},
so that this M2-brane should be 
one of infinitely many M2-branes dissolved into the M5-brane\footnote{See \cite{BLG M5} for the relation to the BLG model.}.
This is consistent with the fact that there must be the M5-brane for the charge conservation\,\cite{ChargeConservation}.
Consequently the NC M5-brane should be regarded as a bound state of 
an M5-brane and M2-branes.

\subsection{Non-commutative M9-branes}\label{sec:NC M9}

We will consider two types of non-commutative M9-branes.
We may choose $\{0,1,\cdots,9\}$ as 
the M9-brane world-volume directions without loss of generality.
The Dirichlet direction is the $10$-th direction which is denoted as $\natural$ below
to avoid confusion.

\subsubsection{Non-commutative M9-brane with an electric flux}
First we consider the following gluing matrix
\begin{align}
M=&\, h_0\Gamma^{01\cdots 9}+h_1\Gamma^{34\cdots 9}~,
\label{M9 glue 1}
\end{align}
which reduces to the gluing matrix for an M9-brane
when  $h_1=0$.
We shall turn on $\CH_{012}$,
and examine
\begin{eqnarray}
\CH_{012}\bar\theta \Gamma^2\xi +\bar\theta\Gamma_{01}\xi=0~.
\label{M9 1}
\end{eqnarray}
Since 
\begin{align}
\bar\theta =\bar\theta M'~,~~~
M'\equiv -h_0\Gamma^{01\cdots 9}+h_1\Gamma^{34\cdots 9}~,
\end{align}
we derive
\begin{align}
\bar\theta\Gamma^2\xi=&\frac{1}{2} \bar\theta(M'\Gamma^2 +\Gamma^2 M)\xi
=-h_0\bar\theta\Gamma^{01\cdots 9}\Gamma^2\xi
=h_0\bar\theta\Gamma^{013\cdots 9}\xi~,\\
\bar\theta\Gamma_{01}\xi=&\frac{1}{2} \bar\theta(M'\Gamma_{01} +\Gamma_{01} M)\xi
=h_1\bar\theta\Gamma^{3\cdots 9}\Gamma_{01}\xi
=-h_1\bar\theta\Gamma^{013\cdots 9}\xi~.
\end{align}
This shows that \bref{M9 1} is satisfied when $\CH_{012}h_0-h_1=0$,
i.e. $\CH_{012}= h_1/h_0$.
The equation \bref{M9 1}
with the replacement $(012)\to(120)$ and
that with the replacement  $(012)\to(201)$
are
treated similarly,
and satisfied when $\CH_{012}= h_1/h_0$.
It is obvious that the latter equation in \bref{BC2}
is satisfied 
\begin{eqnarray}
\bar\theta \Gamma^\natural\xi=\frac{1}{2}\bar\theta(M' \Gamma^\natural+ \Gamma^\natural M)\xi=0~,
\end{eqnarray}
 since $\Gamma^\natural M=-M' \Gamma^\natural$.
As a result,  the gluing matrix \bref{M9 glue 1} with $\CH_{012}= h_1/h_0$
eliminates the BRST surface terms.
For $M^2=h_0^2-h_1^2\equiv 1$, we choose $h_0=\cosh \varphi$ and $h_1=\sinh\varphi$.
In this parametrization, $M$ and $\CH_{012}$ are expressed as
\begin{eqnarray}
M=\e^{\varphi \Gamma^{012}}\Gamma^{01\cdots 9}~,~~~
\CH_{012}=\tanh \varphi~.
\label{NC M9 1}
\end{eqnarray}

When $\varphi =0$, 
it represents a commutative M9-brane characterized by
$M=\Gamma^{01\cdots 9}$ and $\CH_{012}=0$.
On the other hand, when $\varphi\to \infty$,
the boundary condition $\theta =M\theta$ reduces to
$0=(\Gamma^{01\cdots 9} + \Gamma^{3\cdots 9})\theta=\Gamma^{01\cdots 9}(1 - \Gamma^{012})\theta$,
i.e. $\theta = \Gamma^{012}\theta$.
As $\CH_{012}\to 1$, it may represent
an M2-brane
with a critical flux  $\CH_{012}= 1$.
For the charge conservation, there must be another brane
other than  the M2-brane.
We may expect that there should be an M9-brane behind the M2-brane
with a critical flux.
To confirm this expectation we need to know the M9-brane effective action
which describes the coupling to the M2-brane.
Further study is needed to clarify this point.

\subsubsection{Non-commutative M9-brane with a magnetic flux} 

Next,
we shall consider the following gluing matrix
\begin{align}
M=&\, h_0\Gamma^{01\cdots 9}+h_1\Gamma^{01\cdots 6}~,
\label{M9 glue 2}
\end{align}
and turn on $\CH_{789}$\,.
Since 
\begin{align}
\bar\theta =\bar\theta M'~,~~~
M'\equiv -h_0\Gamma^{01\cdots 9}+h_1\Gamma^{01\cdots 6}~,
\end{align}
we derive
\begin{align}
\bar\theta\Gamma^9\xi=&\frac{1}{2} \bar\theta(M'\Gamma^9 +\Gamma^9 M)\xi
=-h_0\bar\theta\Gamma^{01\cdots 9}\Gamma^9\xi
=-h_0\bar\theta\Gamma^{01\cdots 8}\xi~,\\
\bar\theta\Gamma_{78}\xi=&\frac{1}{2} \bar\theta(M'\Gamma_{78} +\Gamma_{78} M)\xi
=h_1\bar\theta\Gamma^{01\cdots 6}\Gamma_{78}\xi
=h_1\bar\theta\Gamma^{01\cdots 8}\xi~.
\end{align}
This shows that
\begin{eqnarray}
\CH_{789}\bar\theta \Gamma^9\xi +\bar\theta\Gamma_{78}\xi=0~
\label{M9 2}
\end{eqnarray}
is satisfied when $-\CH_{789}h_0+h_1=0$,
i.e. $\CH_{789}= h_1/h_0$.
The equation \bref{M9 2} with the replacement $(789)\to(897)$
and that with the replacement $(789)\to(978)$
are shown to be  treated similarly and are satisfied when $\CH_{789}= h_1/h_0$.
It is obvious that the latter equation in \bref{BC2}
is satisfied
 since $\Gamma^\natural M=-M' \Gamma^\natural$.
As a result,  the gluing matrix \bref{M9 glue 2} with $\CH_{789}= h_1/h_0$
eliminates the BRST surface terms.
For $M^2=h_0^2+h_1^2\equiv 1$, we choose $h_0=\cos \varphi$ and $h_1=\sin\varphi$.
In this parametrization, $M$ and $\CH_{789}$ are expressed as
\begin{eqnarray}
M=\e^{\varphi \Gamma^{789}}\Gamma^{01\cdots 9}~,~~~
\CH_{789}=\tan \varphi~.
\label{NC M9 2}
\end{eqnarray}

When $\varphi =0$, 
it represents a commutative M9-brane characterized by
$M=\Gamma^{01\cdots 9}$ and $\CH_{789}=0$.
On the other hand, when $\varphi=\pi/2$,
the gluing matrix reduces to 
 $M=\Gamma^{01\cdots 6}$
 and the flux diverges $\CH_{789}\to \infty$.
It seems that this describes a 6-brane,
but there is no seven-form gauge potential in eleven-dimensions.
In ten-dimensions, however, we have a RR seven-form gauge potential $C_7$
which is dual to a RR one-form gauge potential $C_1$.
The gauge potential $C_7$ couples to a D6-brane which is characterized by a
harmonic function $H$ on the space $\bbE^3$
transverse to the world-volume $\bbE^{1,6}$.
An eleven-dimensional lift of the D6-brane is known as a KK monopole
which is magnetically charged with respect to $C_1$
and takes the form
\cite{KK7M}
\begin{align}
\d s^2=&\d s^2(\bbE^{1,6}) + H(y) \d y^i \d y_i +H(y)^{-1}(\d z+\d y^i C_i(y))^2~,
\nn\\
F_{ij}\equiv &\pa_i C_j(y) - \pa_j C_i(y)
=\epsilon_{ijk} \pa_k H(y)~,
\end{align}
where $y^i~(i=1,2,3)$ are coordinates on $\bbE^3$.
The direct dimensional reduction with respect to $z$
leads to the D6-brane solution.
We will interpret the boundary characterized by $M=\Gamma^{01\cdots 6}$
as the KK monopole.
In the present context, the KK monopole  
extends along $\bbE^{1,6}$ spanned by  $\{0,1,\cdots,6\}$
and $H$ is a harmonic function on $\bbE^3$ spanned by $\{7,8,9\}$.
As $\CH_{789}\to \infty$, 
there should be infinitely many KK monopoles
so that $H(y)$ has infinitely many poles on $\bbE^3$.
These KK monopoles have dissolved inside an M9-brane.
Consequently the NC M9-brane should be regarded as a bound state of 
an M9-brane and KK monopoles.

\subsection{Supersymmetry of non-commutative M-branes}
\label{sec:SUSY}

In this subsection, we shall show that the NC M-branes derived in
the sections \ref{sec:NC M5}
and
\ref{sec:NC M9}
are
half supersymmetric.
For this purpose we will 
examine the surface terms of the supersymmetry transformations
and show that
they are deleted by the boundary conditions examined in the previous subsections.

\medskip

To see that the action \bref{action PS} is invariant under the supersymmetry transformations \bref{SUSY},
we need to perform partial integration.
For an open supermembrane we are considering, the partial integration leaves a surface term
which has to be deleted by an appropriate boundary condition
for some amount of supersymmetry to be preserved.
We find that the surface terms come from the Wess-Zumino term \bref{WZ}
and 
take the form
\begin{align}
 \delta_\epsilon S_\mathrm{pure}|=&\int_{\pa\Sigma}
 \left(
 \CL^{(2)}_\mathrm{SUSY}
 + \CL^{(4)}_\mathrm{SUSY}
 + \CL^{(6)}_\mathrm{SUSY}
 \right)
 \, \d S~,\label{SUSY surface}\\
\CL^{(2)}_\mathrm{SUSY}=&
\frac{i}{4} \epsilon^{ijk}\left(
\CH_{\mu\nu\rho}\bar\theta\Gamma^\rho \epsilon
+\bar\theta \Gamma_{\mu\nu}\epsilon
\right) \pa_j x^\mu \pa_k x^{\nu} n_i~,
 \\
\CL^{(4)}_\mathrm{SUSY}=&
\frac{1}{24} \epsilon^{ijk}\left(
\bar\epsilon \Gamma_{\mu\nu}\theta\cdot \bar\theta\Gamma^\mu \pa_j\theta
+\bar\theta\Gamma_{\mu\nu}\pa_j\theta\cdot \bar\epsilon \Gamma^\mu \theta
\right) \pa_k x^\nu n_i~,
\label{SUSY surface 4}\\
\CL^{(6)}_\mathrm{SUSY}=&
\frac{i}{144} \epsilon^{ijk}\left(
\bar\epsilon \Gamma_{\mu\nu}\theta\cdot \bar\theta\Gamma^\mu \pa_j\theta
+\bar\theta\Gamma_{\mu\nu}\pa_j\theta\cdot \bar\epsilon \Gamma^\mu \theta
\right) \bar\theta \Gamma^\nu \pa_k \theta\,n_i~.
\label{SUSY surface 6}
\end{align}
A derivation of them was given in the appendix B.
Now, we shall compare these surface terms with the BRST surface terms in \bref{L2},\bref{L4}
and \bref{L6}.
We found that
\begin{align}
\CL^{(2)}_\mathrm{SUSY} = \CL^{(2)}|_{\xi=-\epsilon}~,~~~
\CL^{(4)}_\mathrm{SUSY} = \CL^{(4)}|_{\xi=-\epsilon/3}~,~~~
\CL^{(6)}_\mathrm{SUSY} = \CL^{(6)}|_{\xi=-\epsilon/3}~.
\end{align}
It implies that the boundary conditions which eliminate the BRST surface terms
also eliminate the surface terms of supersymmetry\footnote{
For the $\kappa$-symmetric open supermembrane,
see \cite{EMM},
where it was shown that the boundary conditions to eliminate the $\kappa$-symmetry surface term
will preserve a half of the supersymmetries.
}.
As a result, we may conclude that 
the NC M-branes obtained in the sections \ref{sec:NC M5} and \ref{sec:NC M9}
would preserve a half of 32 supersymmetries
so that they are 1/2 BPS objects.

\section{Summary and discussions}

We examined boundary conditions
for the BRST symmetry of the open supermembrane
with a constant flux in the  pure spinor formalism.
The boundary conditions lead to a possible Dirichlet branes
of an open supermembrane.
It is found that the surface terms coincide with
those for the $\kappa$-variation of the $\kappa$-symmetric open supermembrane
examined in \cite{EMM, SY NC M5, SY intersecting, SY NC M5 2}
if we replace $\xi$ for $\delta_\kappa \theta$.
So we have obtained the NC M5-brane
derived there.
In addition to the NC M5-brane,
we found two types of NC M9-branes in this paper.
One is the NC M9-brane with an electric flux
characterized by \bref{NC M9 1}.
It  reduces in the critical electric flux  limit to an M2-brane on an M9-brane.
Another is the NC M9-brane with a magnetic flux
 characterized by \bref{NC M9 2}.
It reduces in the strong flux limit  to infinitely many KK monopoles dissolved into an M9-brane.
It is argued that this NC M9-brane should be regarded as a bound state of 
an M9-brane and KK monopoles.
Furthermore we have examined the surface terms for the supersymmetry transformations
of the open supermembrane action with a constant three-form flux in the pure spinor formalism.
We found that the surface term $\CL_\mathrm{SUSY}^{(n)}$
in \bref{SUSY surface}
for the supersymmetry variation is proportional to the surface term
$\CL^{(n)}$ in \bref{L}
for the BRST variation
if we replace the supersymmetry parameter $\epsilon$ for $\xi$.
Consequently we have concluded  that the NC M-branes
obtained here should preserve a half of 32 supersymmetries
and are 1/2 BPS objects.

\medskip

In this paper, we have examined an open supermembrane in the pure spinor formalism,
instead of the $\kappa$-symmetric open supermembrane.
One of advantages of our approach is to consider the BRST symmetry
which is expected to survive quantum corrections.
We found that
our results are consistent with the previous ones
obtained from the $\kappa$-symmetry arguments.
This implies that our results may give a quantum consistency check for the previous ones.

For the NC M9-brane
with an electric flux  characterized by \bref{NC M9 1}, 
we assumed that there should be an M9-brane behind the M2-brane
in the critical flux limit.
To confirm this point we need to know the M9-brane effective action
which describes the coupling to the M2-brane.
It is interesting for us to clarify this point.

We obtained NC M-branes as solutions of boundary conditions
for an open supermembrane.
They are expected to be obtained also as classical solutions
of M-brane world-volume equations of motion.
Especially
they should preserve a half of 32 supersymmetries
from our analysis in the section \ref{sec:SUSY}.
Solutions of the world-volume equations of motion will help
us to examine properties of the NC M-branes obtained in this paper.
It is also interesting to pursue supergravity solutions of NC M-branes.

%%% added in v2  %%%

It is known that in a general background, the classical BRST invariance of
 an open pure spinor superstring implies that the background fields satisfy full non-linear
 equations of motion for a supersymmetric Born-Infeld action
\cite{PS open background}.
This is the open string version of \cite{PS closed background}
in which  the classical BRST invariance of a closed pure spinor superstring
in a curved background
is shown to imply that the background fields satisfy 
full non-linear equations of motion
for the type-II supergravity.
There are similar results 
for the classical $\kappa$-invariance of an open Green-Schwarz superstring \cite{GS open background} and a closed Green-Schwarz  
superstring \cite{GS closed background}, respectively.
It is interesting to extract non-commutative M-brane equations of motion by requiring BRST invariance in the pure spinor supermembrane with background fields.

%%%%%%%%%%%%

An obvious generalization of our  analysis
is to examine NC M-branes in curved backgrounds,
such as the \adss{4}{7}
background.
In \cite{AdS M-branes}, commutative M-branes are
derived as Dirichlet branes of the $\kappa$-symmetric open supermembrane 
 in \adss{4}{7} and \adss{7}{4}. .
 It is interesting to examine NC M-branes in 
\adss{4}{7}
 by using open supermembrane with constant fluxes
 in the pure spinor  formalism.
 As for D-branes in \adss{5}{5},
we examined them from the BRST symmetry of the pure spinor superstring in \cite{HS}.
Furthermore recently world-sheet supersymmetries are examined in \cite{PS}.
It is also interesting to examine NC D-branes
by using the pure spinor superstring
with two-form fluxes.
We expect that this analysis may support the previous results obtained from the $\kappa$-symmetry
arguments \cite{SY NC AdS}.
We hope to report these issues in the near future.

\section*{Acknowledgments}

The authors would like to thank Takanori Fujiwara, Yoshifumi Hyakutake and Kentaroh Yoshida
 for useful comments.
MS would like to  appreciate the organizers of the conference
``Progress in Quantum Field Theory and String Theory II"
held at Osaka City University, March 27-31, 2017
for their kind hospitality.
There he reported the main results in this paper.
SH would like to thank the Yukawa Institute for Theoretical Physics
at Kyoto University for hospitality during the workshop YITP-W-17-08
``Strings and Fields 2017."
 
\appendix

\section{NC M5-brane}

We will derive the NC M5-branes given in the section \ref{sec:NC M5}.
Consider the gluing matrix of the form
\begin{eqnarray}
M=h_0\Gamma^{\bar\mu_0\cdots\bar\mu_5} +h_1\Gamma^{\bar\mu_0\cdots\bar\mu_2}~.
\end{eqnarray}
We note that this reduces to the gluing matrix for an M5-brane when $h_1=0$.
For $M^2=1$,
we demand $-s_0h_0^2-s_1h_1^2\equiv 1$.
We have introduced $s_0$ and $s_1$
such that
 $s_0=-1$ when $0\in \{\bar\mu_0,\cdots,\bar\mu_5\}$
and $s_0=+1$ otherwise,
while  
$s_1=-1$ when $0\in \{\bar\mu_0,\cdots,\bar\mu_2\}$
and $s_1=+1$ otherwise.
For reality of $\theta$, either $s_0$ or $s_1$ must be $-1$. 
Noting that 
$s_1=1$ follows from $s_0=1$,
we set $s_0=-1$.
Introducing fluxes $\CH_{\bar\mu_0\cdots\bar\mu_2}$
and $\CH^{\bar\mu_3\cdots\bar\mu_5}$
we obtain a non-trivial solution.
Since
\begin{align}
\bar\theta =&\, \theta^{\rm T}M^{\rm T} C =\bar\theta M'~,~~~
M'\equiv -h_0\Gamma^{\bar\mu_0\cdots\bar\mu_5} +h_1\Gamma^{\bar\mu_0\cdots\bar\mu_2}~,
\\
\bar\theta\Gamma_{\bar\mu_0\bar\mu_1}\xi=&\,
\frac{1}{2}\bar\theta(M'\Gamma_{\bar\mu_0\bar\mu_1}+\Gamma_{\bar\mu_0\bar\mu_1} M)\xi=
\frac{1}{2}\bar\theta\Gamma_{\bar\mu_0\bar\mu_1}(M'+M)\xi=
-h_1 \bar\theta\Gamma^{\bar\mu_2}\xi~,
\\
\bar\theta\Gamma_{\bar\mu_5}\xi=&\,
\frac{1}{2}\bar\theta(M'\Gamma_{\bar\mu_5}+\Gamma_{\bar\mu_5} M)\xi=
\frac{1}{2}\bar\theta\Gamma_{\bar\mu_5}(-M'+M)\xi=
-h_0 \bar\theta\Gamma^{\bar\mu_0\cdots\bar\mu_4}\xi~,
\\
\bar\theta\Gamma^{\bar\mu_3\bar\mu_4}\xi=&\,
\frac{1}{2}\bar\theta(M'\Gamma^{\bar\mu_3\bar\mu_4}+\Gamma^{\bar\mu_3\bar\mu_4} M)\xi=
\frac{1}{2}\bar\theta\Gamma^{\bar\mu_3\bar\mu_4}(M'+M)\xi=
h_1 \bar\theta\Gamma^{\bar\mu_0\cdots\bar\mu_4}\xi~,
\\
\bar\theta\Gamma_{\bar\mu_2\bar\mu_3}\xi=&\,
\frac{1}{2}\bar\theta(M'\Gamma_{\bar\mu_2\bar\mu_3}+\Gamma_{\bar\mu_2\bar\mu_3} M)\xi=
\frac{1}{2}\bar\theta\Gamma_{\bar\mu_2\bar\mu_3}(-M+M)\xi=0~,
\end{align}
\bref{BC1}  implies that
\begin{eqnarray}
\CH_{\bar\mu_0\cdots\bar\mu_2}-h_1=0~,~~~
-h_0 \CH^{\bar\mu_3\cdots\bar\mu_5} +h_1=0~.
\end{eqnarray}
Substituting them into  $h_0^2-s_1h_1^2=1$,
we obtain
\begin{eqnarray}
\frac{1}{(\CH_{\bar\mu_0\cdots\bar\mu_2})^2}-\frac{1}{(\CH^{\bar\mu_3\cdots\bar\mu_5} )^2}=-s_1~.
\end{eqnarray}
This is nothing but the self-duality condition
\cite{SD} for the two-form gauge field on the M5-brane world-volume
\cite{SD2}.
It is straightforward to see that  the latter condition in \bref{BC2} is satisfied.

For $s_1=-1$, we may take  $M$ as $M=h_0\Gamma^{012345}+h_1\Gamma^{012}$
without loss of generality.
Parametrizing $h_0$ and $h_1$ as $h_0=\cos\varphi$ and $h_1=\sin\varphi$,
we can express it as $M=\e^{\varphi\Gamma^{345}}\Gamma^{01\cdots5}$.
Fluxes are $\CH_{012}=\sin\varphi$ and $\CH^{345}=\tan\varphi$.
On the other hand for $s_1=1$, we may take  $M$ as $M=h_0\Gamma^{012345}+h_1\Gamma^{345}$
without loss of generality.
Parametrizing $h_0$ and $h_1$ as $h_0=\cosh\varphi$ and $h_1=\sinh\varphi$,
we can express it as $M=\e^{\varphi\Gamma^{012}}\Gamma^{01\cdots5}$.
Fluxes are $\CH_{345}=\sinh\varphi$ and $\CH^{012}=\tanh\varphi$.

\section{Supersymmetry surface term}

We shall derive the supersymmetry surface terms \bref{SUSY surface 4}
and \bref{SUSY surface 6}  in section  \ref{sec:SUSY} below.

The surface term $\CL^{(4)}_\mathrm{SUSY}$
given in \bref{SUSY surface 4}
may be derived from
terms contained in  $ \delta_\epsilon \CL_\mathrm{WZ}$
which are the four-th order terms in fermions
as follows
\begin{align}
\delta_\epsilon \CL_\mathrm{WZ}|_{\theta^3\epsilon}=&
\frac{1}{8} \epsilon^{ijk}\left(\bar\epsilon \Gamma_{\mu\nu} \pa_i\theta\cdot\bar\theta\Gamma^\mu\pa_j \theta
-\bar\theta\Gamma_{\mu\nu}\pa_i\theta\cdot\bar\epsilon \Gamma^\mu \pa_j\theta
\right)\,\pa_k x^\nu
\\
=&
\frac{1}{8} \epsilon^{ijk}\left(\frac{1}{2}\bar\epsilon \Gamma_{\mu\nu} \theta\cdot \pa_i \bar\theta\Gamma^\mu\pa_j \theta
+\frac{1}{2}\pa_i\bar\theta\Gamma_{\mu\nu}\pa_j\theta\cdot\bar\epsilon \Gamma^\mu\theta
\right)\,\pa_k x^\nu
\\=&
-\frac{1}{2} \delta_\epsilon \CL_\mathrm{WZ}|_{\theta^3\epsilon}
+\pa_i\left[
\frac{1}{16}\epsilon^{ijk}
\left(\bar\epsilon \Gamma_{\mu\nu} \theta\cdot \bar\theta\Gamma^\mu\pa_j \theta
+\bar\theta\Gamma_{\mu\nu}\pa_j\theta\cdot\bar\epsilon \Gamma^\mu\theta
\right)\,\pa_k x^\nu
\right]
\\=&
\pa_i\left[
\frac{1}{24}\epsilon^{ijk}
\left(\bar\epsilon \Gamma_{\mu\nu} \theta\cdot \bar\theta\Gamma^\mu\pa_j \theta
+\bar\theta\Gamma_{\mu\nu}\pa_j\theta\cdot\bar\epsilon \Gamma^\mu\theta
\right)\,\pa_k x^\nu
\right]~.
\end{align}
In the second equality we have utilized the Fierz identity,
and for the third equality partial integration has been performed for $\pa_i\bar\theta$.
Similarly, 
examining terms contained in  $ \delta_\epsilon \CL_\mathrm{WZ}$
which are six-th order in fermions,
we derive \bref{SUSY surface 6} as follows.
\begin{align}
\delta_\epsilon \CL_\mathrm{WZ}|_{\theta^5\epsilon}=&
\frac{i}{48} \epsilon^{ijk}\left(\bar\epsilon \Gamma_{\mu\nu} \pa_i\theta\cdot\bar\theta\Gamma^\mu\pa_j \theta
-\bar\theta\Gamma_{\mu\nu}\pa_i\theta\cdot\bar\epsilon \Gamma^\mu \pa_j\theta
\right)\,\bar\theta\Gamma^\nu\pa_k \theta
\\
=&
\frac{i}{48} \epsilon^{ijk}\left(\frac{1}{2}\bar\epsilon \Gamma_{\mu\nu} \theta\cdot \pa_i \bar\theta\Gamma^\mu\pa_j \theta
+\frac{1}{2}\pa_i\bar\theta\Gamma_{\mu\nu}\pa_j\theta\cdot\bar\epsilon \Gamma^\mu\theta
\right)\,\bar\theta\Gamma^\nu\pa_k \theta
\\=&
-\frac{1}{2} \delta_\epsilon \CL_\mathrm{WZ}|_{\theta^5\epsilon}
+\pa_i\left[
\frac{1}{96}\epsilon^{ijk}
\left(\bar\epsilon \Gamma_{\mu\nu} \theta\cdot \bar\theta\Gamma^\mu\pa_j \theta
+\bar\theta\Gamma_{\mu\nu}\pa_j\theta\cdot\bar\epsilon \Gamma^\mu\theta
\right)\,\bar\theta\Gamma^\nu\pa_k \theta
\right]
\nn\\
&+\frac{i}{96}\epsilon^{ijk}\left(
\bar\epsilon\Gamma_{\mu\nu}\theta\cdot\bar\theta \Gamma^\mu \pa_i\theta
+\bar\theta \Gamma_{\mu\nu}\pa_i\theta\cdot \bar\epsilon\Gamma^\mu \theta
\right)\pa_j\bar\theta\Gamma^\nu\pa_k \theta
\label{SUSY L6}
\\=&
\pa_i\left[
\frac{1}{144}\epsilon^{ijk}
\left(\bar\epsilon \Gamma_{\mu\nu} \theta\cdot \bar\theta\Gamma^\mu\pa_j \theta
+\bar\theta\Gamma_{\mu\nu}\pa_j\theta\cdot\bar\epsilon \Gamma^\mu\theta
\right)\,\bar\theta\Gamma^\nu\pa_k \theta
\right]~.
\end{align}
In the second equality we have utilized the Fierz identity,
and for the third equality partial integration has been performed for $\pa_i\bar\theta$.
The last line in \bref{SUSY L6}
is eliminated by the Fierz identity.

\end{document}